\documentstyle[prc,aps,preprint]{revtex}
\begin{document} \draft
\date{\today}
\title{The $\phi a_0\gamma$- and $\phi\sigma\gamma$-vertices in light cone QCD }

\author{A. Gokalp~\thanks{agokalp@metu.edu.tr} and
        O. Yilmaz~\thanks{oyilmaz@metu.edu.tr}}
\address{ {\it Physics Department, Middle East Technical University,
06531 Ankara, Turkey}}
\maketitle

\begin{abstract}
We study $\phi a_0\gamma$- and $\phi\sigma\gamma$-vertices in the
framework of the light cone QCD sum rules and we estimate the
coupling constants g$_{\phi a_0\gamma}$ and g$_{\phi\sigma\gamma}$
utilizing $\omega\phi$-mixing. We compare our results with the
previous estimations of these coupling constants in the literature
obtained from phenomenological considerations.
\end{abstract}

\thispagestyle{empty} ~~~~\\ \pacs{PACS numbers:
12.38.Lg;13.40.Hq;14.40.Aq }
\newpage
\setcounter{page}{1}

The studies of $\phi(1020)$ meson and in particular its radiative
decays have been important sources of information in hadron
physics. The Novosibirsk CMD \cite{R1} and SND \cite{R2}
collaborations have reported measurements of radiative
$\phi\rightarrow\pi^0\pi^0(\eta)\gamma$ and
$\phi\rightarrow\pi^+\pi^-\gamma$ decays, and the high luminosity
Frascati $\phi$-factory $DA\Phi NE$ will soon be performing
precise measurements of radiative $\phi$ decays. These information
will help us to increase our understansing of the complicated
dynamics of meson physics in the 1 GeV energy region. In this
energy region, low-mass scalar mesons may also play an important
role. The incorporation of the role of the scalar resonances in
processes involving vector mesons provides an opportunity  to
increase our insight of the dynamics of the meson physics. Among
the processes involving the vector and scalar mesons, the $\phi
a_0\gamma$- and $\phi\sigma\gamma$-vertices are interesting and
important for several reasons. The $\phi a_0\gamma$-vertex plays a
role in the study of the radiative
$\phi\rightarrow\pi^0\eta\gamma$ decay \cite{R3}, and the
knowledge of the $\phi\sigma\gamma$-vertex is needed in the
analysis of the decay mechanism of the
$\phi\rightarrow\pi^0\pi^0\gamma$ decay \cite{R4}. Furthermore,
the coupling constant g$_{\phi\sigma\gamma}$ is needed in the
study of the structure of the $\phi$ meson photoproduction
amplitude on nucleons in the near threshold region based on the
one-meson exchange and Pomeron-exchange mechanisms \cite{R5}. In
the present work, we employ the light cone QCD sum rules method to
study the $\phi a_0\gamma$- and $\phi\sigma\gamma$-vertices, and
utilizing $\omega\phi$-mixing we estimate the g$_{\phi a_0\gamma}$
and g$_{\phi\sigma\gamma}$ coupling constants.

In order to analyze the $\phi a_0\gamma$- and
$\phi\sigma\gamma$-vertices using light cone QCD sum rules, we
begin with the observation that the $\phi\rightarrow\pi^0\gamma$
decay width vanishes if the $\phi$ meson is a pure $s\overline{s}$
state. The measured width
$\Gamma(\phi\rightarrow\pi^0\gamma)=(5.6\pm 0.5)$ KeV \cite{R6} is
significantly different from zero which is explained as primarily
being  due to $\omega-\phi$ mixing. Bramon et al. \cite{R7} have
recently studied radiative $VP\gamma$ transitions between vector
(V) mesons and pseudoscalar (P) mesons, and using the available
experimental information they have determined the mixing angle for
$\omega-\phi$ as well as other relevant parameters of
$\omega-\phi$ and $\eta-\eta^\prime$ systems. In this work, we
follow their treatment and we write the physical $\omega$ and
$\phi$ meson states as
\begin{eqnarray}\label{e1}
  \mid \omega>&=&\cos\theta_v\mid\omega_0>-\sin\theta_v\mid\phi_0> \nonumber \\
  \mid \phi>&=&\sin\theta_v\mid\omega_0>+\cos\theta_v\mid\phi_0>~~,
\end{eqnarray}
where $\mid\omega_0>=\frac{1}{\sqrt{2}}\mid
u\overline{u}+d\overline{d}>$ and $\mid\phi_0>=\mid
s\overline{s}>$ are the non-strange and the strange basis states.
The mixing angle has been determined from the available
experimental data by Bramon et al. as $\theta=(3.4\pm 0.2)^o$
\cite{R7}. We, therefore, choose the interpolating currents for
$\omega$ and $\phi$ mesons defined in the quark flavour basis as
\begin{eqnarray}\label{e2}
  j_\mu^\omega&=&\cos\theta_v j_\mu^{\omega_0}-\sin\theta_v j_\mu^{\phi_0} \nonumber \\
  j_\mu^\phi&=&\sin\theta_v j_\mu^{\omega_0}+\cos\theta_v j_\mu^{\phi_0}~~,
\end{eqnarray}
where $j_\mu^{\omega_0}=\frac{1}{6}( \overline{u}\gamma_\mu u
+\overline{d}\gamma_\mu d)$ and $j_\mu^{\phi_0}=-\frac{1}{3}
\overline{s}\gamma_\mu s$ \cite{R8}.

In order to study the $\phi s\gamma$-vertex and to estimate the
coupling constant g$_{\phi s\gamma}$ with s denoting $a_0$ or
$\sigma$ meson, we consider the two point correlation function
with photon
\begin{equation}\label{e3}
  T_{\mu}(p,q)=i\int d^{4}x e^{ip\cdot x}
  <\gamma(q)|T\{j_\mu^\phi(0)j_{s}(x)\}|0>
\end{equation}
where $j_\mu^{\phi}=\sin\theta_v j_\mu^{\omega_0}+\cos\theta_v
j_\mu^{\phi_0}$ and $j_s=\frac{1}{2}(
\overline{u}u+(-1)^I\overline{d}d)$ are the interpolating currents
for $\phi$ and for isoscalar I=0 $\sigma$ meson, and for isovector
I=1 $a_0$ meson. The overlap amplitudes of these interpolating
currents with the meson states are defined as
\begin{eqnarray}\label{e4}
<0|j_{\mu}^\phi|\phi>&=&\lambda_\phi u_\mu \nonumber \\
<0|j_{s}|s>&=&\lambda_s
\end{eqnarray}
where $u_\mu$ is the polarization vector of $\phi$ meson and s
denotes $\sigma$ or $a_0$ meson. The $e^+e^-$ leptonic decay width
of $\phi$ meson neglecting electron mass can  be written as
$\Gamma(\phi\rightarrow
e^+e^-)=\frac{4\pi\alpha^2}{3m_\phi^3}\lambda_\phi^2$. We use the
experimental value for the  branching ratio $B(\phi\rightarrow
e^+e^-)=(2.91\pm 0.07) \times 10^{-4}$ of $\phi$ meson \cite{R6},
and this way we determine the overlap amplitude $\lambda_\phi$ of
$\phi$ meson as $\lambda_\phi=(0.079\pm 0.016)$ GeV$^2$. We have
employed the QCD sum rules method in previous works and we
determined the overlap amplitudes $\lambda_\sigma$ and
$\lambda_{a_0}$ as $\lambda_\sigma=(0.12\pm 0.03)~~GeV^2$
\cite{R9} and $\lambda_{a_0}=(0.21\pm 0.005)~~GeV^2$ \cite{R10},
since they are not available experimentally.

The theoretical part of the sum rule for the coupling constants
g$_{\phi s\gamma}$ is obtained in terms of QCD degrees of freedom
by calculating the two point correlator in the deep Euclidean
region where $p^2$ and $(p+q)^2$ are large and negative. In this
calculation we use the full light quark propagator with both
perturbative and nonperturbative contributions, and it is given as
\cite{R11}
\begin{eqnarray}\label{e5}
  iS(x,0)&=&<0|T\{\overline{q}(x)q(0)\}|0>\nonumber \\
         &=&i\frac{\not x}{2\pi^2x^4}-\frac{<\overline{q}q>}{12}-
         \frac{x^2}{192}m_0^2<\overline{q}q>\nonumber \\
         &~&-ig_s\frac{1}{16\pi^2}\int_0^1du\left\{
         \frac{\not{x}}{x^2}\sigma_{\mu\nu}G^{\mu\nu}(ux)
         -4iu\frac{x_\mu}{x^2}G^{\mu\nu}(ux)\gamma_\nu\right\}+...
\end{eqnarray}
where terms proportional to light quark mass m$_u$ or m$_d$ are
neglected. We note that it is the $j_\mu^{\omega_0}$ part of the
$\phi$ meson interpolating current, that is
$j_\mu^{\phi}=\sin\theta_v j_\mu^{\omega_0}=(1/6)\sin\theta_v
(\overline{u}\gamma_\mu u+\overline{d}\gamma_\mu d)$, that makes a
contribution in the calculation of the theoretical part of the sum
rule. After a straightforward computation we obtain
 \begin{eqnarray}\label{e6}
T_{\mu}(p,q)=4i\int d^4xe^{ipx}A(x_\sigma g_{\mu\tau}-x_\tau
g_{\mu\sigma})<\gamma(q)\mid
\overline{q}(x)\sigma_{\tau\sigma}q(0)\mid 0>
\end{eqnarray}
where $A=\frac{i}{2\pi^2x^4}$. In order to evaluate the two point
correlation function further, we need the matrix elements
$<\gamma(q)|\overline{q}\sigma_{\alpha\beta}q|0>$. These matrix
elements are defined in terms of the photon wave functions
\cite{R12,R13,R14}
\begin{eqnarray}\label{e7}
<\gamma(q)|\overline{q}\sigma_{\alpha\beta}q|0>
&=&ie_q<\overline{q}q> \int_0^1due^{iuqx}\{(\epsilon_\alpha
q_\beta-\epsilon_\beta q_\alpha)\left[\chi\phi(u)+x^2[
g_1(u)-g_2(u)]\right]  \nonumber\\ && + \left[q\cdot
x(\epsilon_\alpha x_\beta-\epsilon_\beta x_\alpha)+\epsilon\cdot
x(x_\alpha q_\beta-x_\beta q_\alpha)\right]g_2(u)\}~~,
\end{eqnarray}
where the parameter $\chi$ is the magnetic susceptibility of the
quark condensate and $e_q$ is the quark charge, $\phi(u)$ stands
for the leading twist-2 photon wave function, while $g_1(u)$ and
$g_2(u)$ are the two-particle photon wave functions of twist-4. In
the further analysis the path ordered gauge factor is omitted
since we work in the fixed point gauge \cite{R15}.

In order to construct the phenomenological part of the two point
function in Eq. 3, we note that the two point function
T$_\mu(p,q)$ satisfies a dispersion relation, and we saturate this
dispersion relation by inserting a complete set of one hadron
states into the correlation function. This way we construct the
phenomenological part of the two point correlation function as
\begin{equation}\label{e8}
  T_{\mu}(p,q)=\frac{<s\gamma|\phi>
  <\phi|j_\mu^\phi|0><0|j_{s}|s>}
  {(p^2-m^2_\phi)({p^\prime}^2-m^2_s)}+...
\end{equation}
where we denote  the contributions from the higher states and the
continuum starting from some threshold $s_0$  by dots. The
coupling constant g$_{\phi s\gamma}$ is defined through the
effective Lagrangian
\begin{equation}\label{e9}
{\cal L}=\frac{e}{m_{\phi}}g_{\phi s\gamma}\partial^\alpha
\phi^\beta(\partial_\alpha A_\beta-\partial_\beta A_\alpha )s
\end{equation}
describing the $\phi s\gamma$-vertex \cite{R16}. The
$<s\gamma|\phi>$ matrix can therefore be written as
\begin{equation}\label{e10}
<s(p^\prime)\gamma(q)\mid\phi(p)>= i\frac{e}{m_\phi}g_{\phi s
\gamma}K(q^2)(p\cdot q ~u_\mu -u\cdot q~ p_\mu)
\end{equation}
where $q=p-p^\prime$ and $K(q^2)$ is a form factor with K(0)=1. In
order to take the contributions coming from the higher states and
the continuum into account, we invoke the quark-hadron duality
prescription and replace the hadron spectral density with the
spectral density calculated in QCD. In accordance with the QCD sum
rules method strategy, we equate the two representations of the
two point correlation function, theoretical and phenomenological,
and we construct the sum rule for the coupling constant g$_\phi
s\gamma$. After evaluating the Fourier transform and then
performing the double Borel transformation with respect to the
variables $Q_1^2=-{p^\prime}^2$ and $Q_2^2=-(p^\prime+q)^2$, we
finally obtain the following sum rule for the coupling constant
g$_{\phi s\gamma}$
\begin{eqnarray}\label{e11}
g_{\phi s\gamma}=\frac{1}{6}\frac{m_\phi
(e_u+(-1)^Ie_d)<\overline{u}u>}{\lambda_\phi\lambda_s}
  e^{m_s^2/M_1^2}e^{m_\phi^2/M_2^2}
 \left\{-M^2\chi\phi(u_0)f_0(s_0/M^2)+4g_1(u_0)\right\} \sin\theta
\end{eqnarray}
where the function $f_0(s_0/M^2)=1-e^{-s_0/M^2}$ is the factor
used to subtract the continuum, $s_0$ being the continuum
threshold, and
\begin{equation}\label{e12}
  u_0=\frac{M_1^2}{M_1^2+M_2^2}~~~~~~,
  M^2=\frac{M_1^2M_2^2}{M_1^2+M_2^2}\nonumber
\end{equation}
with M$_1^2$ and M$_2^2$ are the Borel parameters.

For the numerical evaluation of the sum rules, we use the value
$<\overline{u}u>=(-0.014\pm 0.002)~~GeV^3$ \cite{R8} for the
vacuum condensate, and $\chi=-4.4~~GeV^{-2}$ \cite{R13,R17} for
the magnetic susceptibility of the quark condensate. The leading
twist-2 photon wave function is given as $\phi(u)=6u(1-u)$ and the
two-particle photon wave function of twist-4 is given by the
expression $g_1(u)=-(1/8)(1-u)(3-u)$  \cite{R13}. We use for the
overlap amplitudes the values $\lambda_\sigma=(0.12\pm
0.03)~~GeV^2$, $\lambda_{a_0}=(0.21\pm 0.05)~~GeV^2$, and
$\lambda_\phi=(0.079\pm 0.016)~~GeV^2$ as we have discussed
previously, and we use $m_\phi=1.02~~GeV$,  $m_\sigma=0.5~~GeV$,
and $m_{a_0}=0.98~~GeV$. In order to analyze the dependence of the
coupling constants g$_{\phi\sigma\gamma}$ and g$_{\phi a_0\gamma}$
on the Borel parameters $M_1^2$ and $M_2^2$, we study independent
variations of $M_1^2$ and $M_2^2$. We find that the sum rule for
the coupling constant g$_{\phi\sigma\gamma}$ is quite stable for
$M_1^2=1.2~~GeV^2$ and for $1.0 ~~GeV^2<M_2^2<1.4 ~~GeV^2$, and
that for the coupling constant g$_{\phi a_0\gamma}$ for the values
$M_2^2=2.0~~GeV^2$ and for $1.0 ~~GeV^2<M_2^2<1.4 ~~GeV^2$. We
note that these limits on $M_2^2$ are within the allowed interval
for the vector channel \cite{R18}. Moreover, we study the
dependence of the sum rules on the threshold parameter $s_0$. The
variation of the coupling constants g$_{\phi a_0\gamma}$ and
g$_{\phi\sigma\gamma}$ as a function of the Borel parameter
$M_2^2$ for the values of $s_0=1.5, 1.6, 1.7~~GeV^2$ with
$M_1^2=2.0~~GeV^2$ for g$_{\phi a_0\gamma}$, and for the values of
$s_0=1.1, 1.2, 1.3~~GeV^2$ with $M_1^2=1.2~~GeV^2$ in the case of
g$_{\phi \sigma\gamma}$ are shown in Fig. 1 and in Fig. 2,
respectively, from which we conclude that the variations are quite
stable. The sources contributing to the uncertainties are those
due to variations in the Borel parameters $M_1^2$, $M_2^2$, in the
threshold parameter $s_0$, and in the estimated values of the
vacuum condensate and the magnetic susceptibility. If we take
these uncertainties into account by a conservative estimate, we
obtain the coupling constants g$_{\phi a_0\gamma}$ and
g$_{\phi\sigma\gamma}$ as g$_{\phi a_0\gamma}=(0.11\pm 0.03)$ and
g$_{\phi\sigma\gamma}=(0.036\pm 0.008)$.

In a previous work \cite{R4}, we studied the radiative
$\phi\rightarrow\pi^0\pi^0\gamma$ decay. We considered $\rho$-pole
vector meson dominance amplitude as well as scalar $\sigma$-pole
and $f_0$-pole amplitudes, and by employing the experimental value
for this decay rate and by analyzing the interference effects
between different contributions in the experimental $\pi^0\pi^0$
invariant mass spectrum for the decay
$\phi\rightarrow\pi^0\pi^0\gamma$, we estimated the coupling
constant g$_{\phi\sigma\gamma}$ as g$_{\phi\sigma\gamma}=(0.025\pm
0.009)$ which is in reasonable agreement with our present
calculation utilizing light cone QCD sum rules method. On the
other hand, Friman and Soyeur \cite{R16} in their study of the
photoproduction of $\rho^0$ mesons on proton targets near
threshold showed that photoproduction cross section is given
mainly by $\sigma$-exchange. They calculated the
$\rho\sigma\gamma$-vertex assuming vector meson dominance of the
electromagnetic current and then they performed a fit to the
experimental $\rho^0$ photoproduction data. Their result when
described using an effective Lagrangian for the
$\rho\sigma\gamma$-vertex gave the value
g$_{\rho\sigma\gamma}=2.71$ for this coupling constant. In their
study of the structure of the $\phi$ meson photoproduction
amplitude on nucleons near threshold based on the one-meson
exchange and Pomeron-exchange mechanisms, Titov et al. \cite{R5}
used this value of the coupling constant g$_{\rho\sigma\gamma}$ to
calculate the coupling constants g$_{\phi\sigma\gamma}$ and
g$_{\phi a_0\gamma}$ by invoking unitary symmetry arguments. They
assumed that $\sigma$, $f_0$, and $a_0$ are members of a unitary
nonet, and they obtained the results g$_{\phi\sigma\gamma}=0.047$
and $\mid g_{\phi a_0\gamma}\mid=0.16$ for these coupling
constants. Our results g$_{\phi\sigma\gamma}=(0.036\pm 0.008)$ and
g$_{\phi a_0\gamma}=(0.11\pm 0.03)$ are in agreement with the
values of these coupling constants calculated by Titov et al. and
used in their analysis. However, it should be noted that in our
work we do not make any assumption about the assignment of scalar
states into various unitary nonets, which is not without problems.


\newpage

{\bf Figure Caption:}

\begin{description}
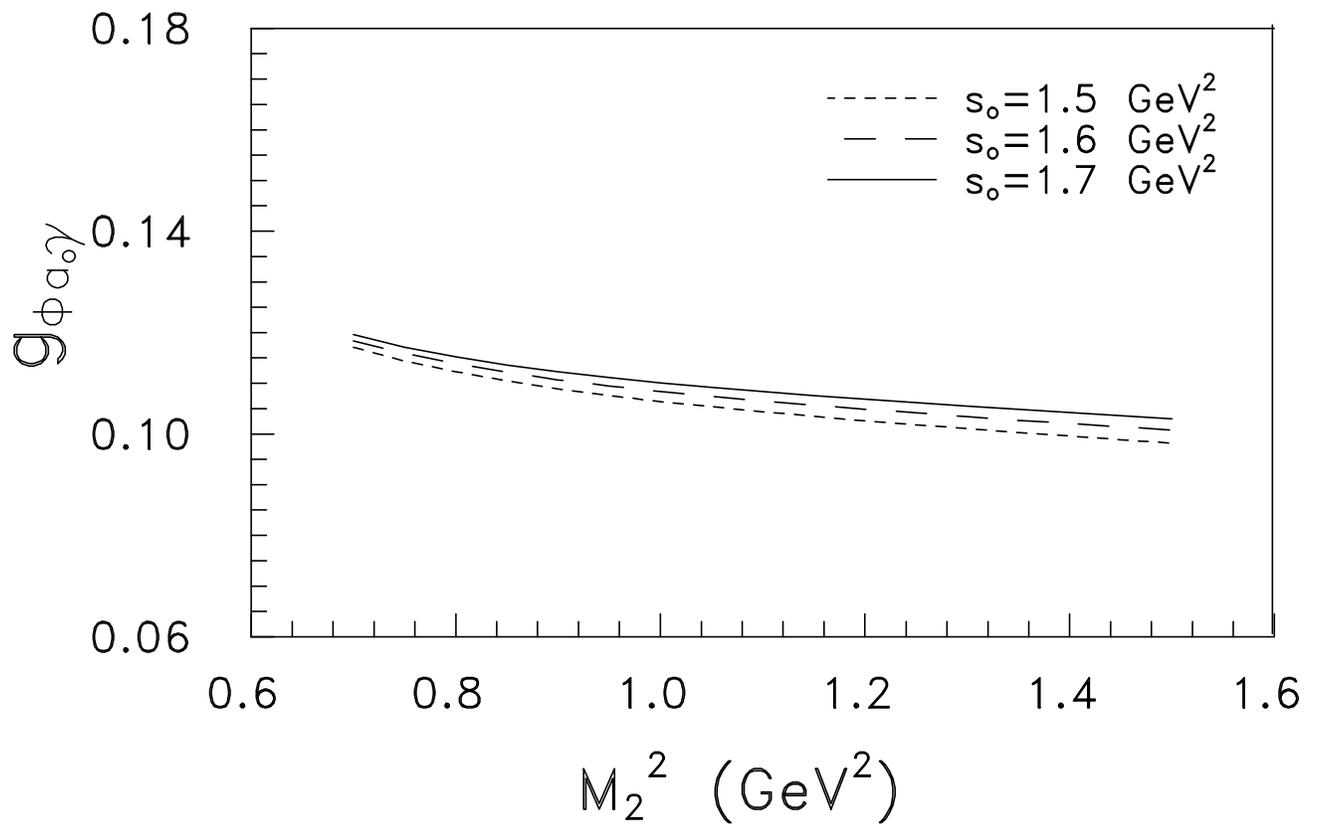
\item[{\bf Figure 1}:] The coupling constant $g_{\phi a_0\gamma}$
as a function of the Borel parameter $M_2^2$ for different values
of the threshold parameters $s_0$ with $M_1^2$=2.0 GeV$^2$.
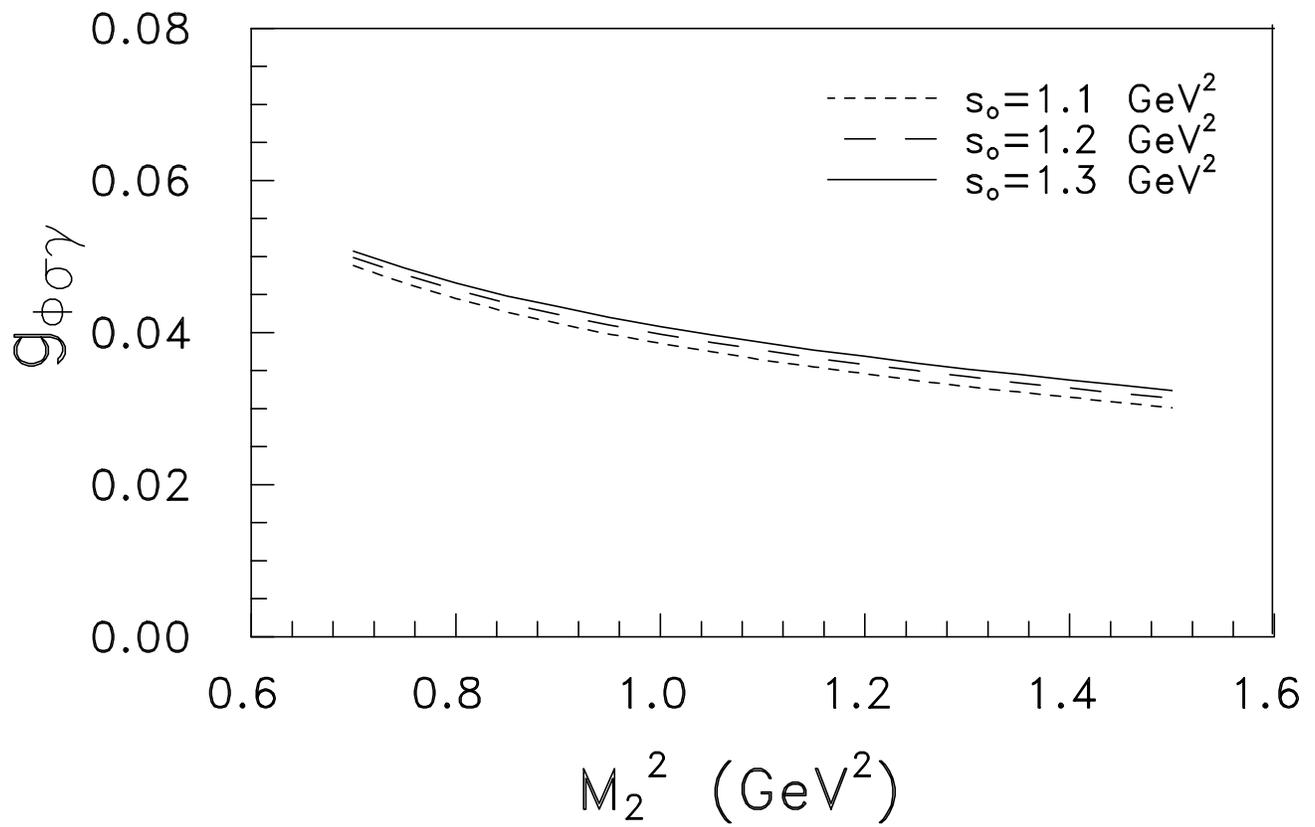
\item[{\bf Figure 2}:] The coupling constant $g_{\phi\sigma\gamma}$
as a function of the Borel parameter $M_2^2$ for different values
of the threshold parameters $s_0$ with $M_1^2$=1.2 GeV$^2$.
\end{description}

\end{document}